\documentclass[preprint2]{aastex63}

\usepackage{amsmath}

\begin{document}
\title{Explaining the cuspy dark matter halo by Landau-Ginzburg theory}
\date{\today}
\author{Dong-Biao Kang}
\email{dbkang@aynu.edu.cn}
\affiliation{Anyang Normal University, 455000, Henan, P. R. China}


\begin{abstract}
Twenty-three years ago, cosmological $N$-body simulations revealed quasi-universal NFW dark matter halos, whose physical origin is still unclear. This work tries to solve this issue by equilibrium statistical mechanics with the Landau-Ginzburg (LG) theory. Firstly, we will introduce this theory in two approaches and reveal the relations between these two approaches' results. Then we replace the order parameter in LG theory by density and consider the dark matter halos as fluctuations from the equilibrium state of the background, and the universal $r^{-1}$ cusp of the equilibrium dark matte halo can be easily obtained. The background can be the main halos for the subhalos or the whole Universe for the main halos. For the cluster halos our work is also consistent with the behaviour of the power spectrum near the cluster scale. This paper suggests that the $r^{-1}$ cusp may originate from the dependence of the fluctuations of the Helmholtz free energy mainly on the density and the density's gradient.
\end{abstract}

\keywords{Cosmology: theory -- dark matter -- large-scale structure of Universe  -- methods: analytical}

\section{Introduction}
\label{sec:intro}
Cosmological simulations have revealed many almost universal properties of the 'isolated' equilibrium cold dark matter halos \citep{nfw97, navarro10}, and the most prominent one may be the NFW density profile
which shows the inner density slope -1 and the outer slope -3. Very recently, \citet{wang19} in their simulations shows this universality can extend to the halos over twenty orders of magnitude in mass.
Because the halos have finite mass, the outer slope should be smaller than -3, while outside the virial radius there is a bump caused by the halos' neighbour in the density profile \citep{prada06}, which indicate that the outer slope should be shallower than some value, so it may be trivial that the outer slope is about -3, so most of the attentions have been paid to the inner slope. By its universality, any explanations invoking the initial conditions may be not convincing enough, so the method of statistical mechanics is always used trying to find the common physical origin of the cusp, such as that \citet{hjorth10} assumes the microstates should be counted in energy space and modifies the stirling approximation in statistical mechanics, which can show the $r^{-1}$ cusp, however, \citet{destri18} finds that \citet{hjorth10}'s model can also allow the existence of the central core; \citet{Pontzen13} believes that, in the canonical ensemble there exists another constraint caused by incomplete relaxation, but it predicts much less low angular momentum orbits than simulations,
although the statistical mechanics for self-gravitating systems still faces some challenges such as the in-equivalence of ensembles, the broken of ergodicity, the thermodynamical limit problem, etc (a recent review in \citet{levin14}), and although there are still other schemes \citep{Dalal10,hansen12,Dekel03, hansen17}. Moreover, the new developments in studies of the vector resonant relaxation and isotropic-nematic phase transition \citep{kocsis15,roupas17,roupas19} also indicate the great potential of statistical mechanics for self-gravitating systems.

In this paper, the almost universal $r^{-1}$ cusp will be explained by the LG theory, which is always used to study the system's long-range correlation of fluctuations from the equilibrium state in the canonical ensemble \citep{michael99}, such as these studies of the density profile in the vapor-liquid interface, the LG coherence length in superconductivity, and others \citep{stefan18,Paoluzzi18}. The structure of the content is as follows: in the next section, we will briefly review the result of the LG theory in two approaches; in section 3, we will apply this theory for simulated dark matter halos; finally£¬we will make some discussions about this method's availability and conclude.

\section{Landau-Ginzburg theory}
\label{sec:basic}
In statistical physics, LG theory describes the long-range correlation of fluctuations from the equilibrium state in an approximate fashion:
\begin{equation}
C\left(\textbf{r}, \textbf{r}^{\prime}\right) \equiv \overline{[\rho(\textbf{r})-\bar{\rho}(\textbf{r})]\left[\rho\left(\textbf{r}^{\prime}\right)-\bar{\rho}\left(\textbf{r}^{\prime}\right)\right]}
\end{equation}
where $C\left(r, r^{\prime}\right)=C\left(\left|\textbf{r}-\textbf{r}^{\prime}\right|\right)$, $\rho(\textbf{r})$ commonly is the order parameter at $\textbf{r}$, which can be the magnetization in Ising model for magnetic materials (This may be the reason why the order parameter was originally denoted by $'m'$ in \citet{michael99}), the local density in liquid-vapour interface, the wave function in BCS theory for superconductivity£¬etc. $\bar{\rho}(\textbf{r})$ denotes the ensemble average (equilibrium) value. To systems with short-range interactions,
\begin{equation}
\label{short-range}
\bar{\rho}(\textbf{r})=\bar{\rho}\left(\textbf{r}^{\prime}\right)=\bar{\rho}(0) \equiv \bar{\rho}.
\end{equation}
\begin{equation}
C(r)=\overline{[\rho(\textbf{r})-\bar{\rho}][\rho(0)-\bar{\rho}]}
\end{equation}
In this paper, we do not consider the effects of fluctuating temperature. With fixed temperature and volume, Landau-Ginzburg theory assumes that the fluctuation of Helmholtz free energy is
\begin{equation}
\label{freeenery}
\Delta F=F-\bar{F}=\int \{a_{1}(\rho-\bar{\rho})+\frac{a_{2}}{2}(\rho-\bar{\rho})^{2}+\frac{b}{2}(\nabla \rho)^{2}\}\mathrm{d}^3 r.
\end{equation}
To ensure that the system is stable at the equilibrium state $\rho=\bar{\rho}$ and $\delta\rho=0$, $a_{2}$ and $b$ are required to be positive. The last term in (\ref{freeenery}) shows $\Delta F$ may be partly contributed by the gradient of $\rho$. From Appendix it can be known that
\begin{equation}
\label{correlationlength}
C(r)\propto \frac{e^{-r/\xi}}{r},
\end{equation}
where
\begin{equation}
\xi=\sqrt{\frac{b}{a_2}}
\end{equation}
is the correlation length. Above results indicate that the two-point correlation function $C(r)\propto 1/r$ for $r\ll\xi$, while $C(r)\propto e^\frac{-r}{\xi}$ for $r\gg\xi$.

The LG theory can be shown in another way \citep{michael99}: the Helmholtz free energy $F$ is the Legendre transformation of the Gibbs free energy $G$
\begin{equation}
\label{func-free}
F(\rho, T)=G+h \rho
\end{equation}
with
\begin{equation}
\label{fluc-frenergy}
dF=-SdT+hdM
\end{equation}
where $T$ is the temperature and $M=\int d^{3} r \rho(\mathbf{r})$. $h$ is the external field, which causes the fluctuations of $M$, so $(h,M)$ is a pair of generalized force and coordinate. (\ref{fluc-frenergy}) shows the fluctuations of Helmholtz free energy come from the fluctuations of the temperature $T$ and order parameter $M$. Then from (\ref{func-free}) and (\ref{fluc-frenergy}),
\begin{equation}
\label{perturbation}
h(\mathbf{r})=\frac{\delta F}{\delta \phi(\mathbf{r})}=a_2 \phi(\mathbf{r})-b \nabla^{2} \phi(\mathbf{r})
\end{equation}
where the last term is obtained by integration by parts and demanding $\delta \rho=0$ at the surface, and $\phi(r)=\rho(r)-\overline{\rho}$. Eq.(\ref{perturbation})  describes the change of order parameter $\rho$ with $r$. If $h$ is assumed to happen at $r=0$ and be localized, i.e.
\begin{equation}
\label{h(r)}
h(\textbf{r})=h_0\delta(\textbf{r})
\end{equation}
 where $\delta(\textbf{r})$ is the Dirac function,  (\ref{perturbation})'s analytical solution in spherical coordinate is
\begin{equation}
\label{phi}
\phi(r)=\rho(r)-\overline{\rho}=\frac{h_{0}}{4 \pi b} \frac{e^{-r / \xi}}{r}.
\end{equation}
which is one of the general solutions of the (\ref{perturbation})'s corresponding homogeneous equation (the other is $e^{r/\xi}/r$, which has been abandoned). (\ref{phi}) is 'coincidentally' proportional to (\ref{correlationlength}), and this 'coincidence' can be explained by the following: including a term
\begin{equation}
-\int d^{3} r \rho(\mathbf{r}) h(\mathbf{r})
\end{equation}
in the Hamiltonian, we have
\begin{equation}
 \overline{\rho(\mathbf{r})}=\frac{\texttt{Tr} \rho(\mathbf{r}) \exp \left\{-\beta\left[H_{0}-\int \mathrm{d}^{3} r^{\prime} h\left(\mathbf{r}^{\prime}\right) \rho\left(\mathbf{r}^{\prime}\right)\right]\right\}}{\textrm{Tr} \exp \left\{-\beta\left[H_{0}-\int \mathrm{d}^{3} r^{\prime} h\left(\mathbf{r}^{\prime}\right) \rho\left(\mathbf{r}^{\prime}\right)\right]\right\}}
\end{equation}
where $\beta$ is a parameter related to temperature and (\ref{h(r)}) still holds, so
\begin{equation}
\label{rho_correlation}
\begin{aligned}
\frac{\delta \overline{\rho(\mathbf{r})}}{\delta h(0)}&=\phi(\mathbf{r}) / h_{0}=\beta\overline{\rho(\mathbf{r}) \rho(0)}- \beta\overline{\rho(\mathbf{r})}~\overline{\rho(0)}\\
                                                      &=\beta\overline{[\rho(\mathbf{r})-\bar{\rho}][\rho(0)-\bar{\rho}]}\propto C(r)\\
\end{aligned}
\end{equation}
where the last second equality uses (\ref{short-range}).

Now we will discuss the scope of application of the two approaches' results. In the first approach we find that the correlation function (\ref{correlationlength}) seems to be regardless of assumptions of (\ref{freeenery}) but requires the distribution of the fluctuations of the free energy $\Delta F$ to satisfy the canonical distribution; while the second approach seems to suggest that the density  (\ref{correlationlength}) only can set up with special case (\ref{h(r)}) but regardless of the distribution of $\Delta F$. However, a simple analysis suggests that if $h(r)$ is shallower than -4 or $h(r)=\delta(x)q(x)$ where $q$ is an arbitrary function of $x$, then $\rho(r)\propto 1/r$ near $r=0$. After numerical check we find that under the above conditions even (\ref{phi}) can be applicable, so (\ref{phi}) may hold with many cases of $h(r)$ regardless of the distribution of $\Delta F$. (\ref{rho_correlation}) connects the density with correlation function, but it requires both localized $h(r)$(\ref{h(r)}) and canonical distribution of $\Delta F$ (\ref{dstri-deltaF}) to set up.
\section{applications for dark matter halos}
 We try to use the LG theory to study the structure of dark matter halos in simulations. We first study these subhalos with scales equal or smaller than the galaxy, and their main halos are equal or smaller than galaxy cluster. The timescale of violent relaxation (\cite{lb67}) is
\begin{equation}
t_v\thicksim(\frac{3 \pi} { 32 G \bar{\rho}})^{1 / 2}.
\end{equation}
During the relaxation all the halos will evolve mutually, and smaller scale halos generally have higher mean density and finish the violent relaxation earlier than larger-scale halos. After enough time the subhalos and main halos generally will reach the quasi-stationary state one after another. The density profile of the main halos $\overline{\rho(r)}$ in this paper is assumed to be well determined by equilibrium statistical mechanics of self-gravitating systems, such as these described by \citet{hjorth10} or \citet{kang11}. Note that in this part we only study the density profile, and these equilibrium density profiles' corresponding free energies may not satisfy the canonical distribution, which, however, does not influence our study by the above section. These subhalos will be treated as the fluctuations from the equilibrium state of the main halos, and the order parameter in LG theory will be the density here. According to the above section, the subhalo's spherical density distribution should satisfy
\begin{equation}
\label{perturbation2}
h(r,r')=a_2 (\rho(r,r')-\overline{\rho(r)})-b \nabla^{2} (\rho(r,r')-\overline{\rho(r)})
\end{equation}
where $r'$ is the central position of subhalos in spherical coordinate $r$ of galaxy cluster halos. We assume that $h(r,r')=h(r-r')=\delta(r-r')g(r-r')$ or $h(r-r')$ is shallower than -4 near $r=r'$, and this assumption may be enough to describe the perturbations such as galaxy merging, gravity instability, etc. Then (\ref{perturbation2})'s solution near $r=r'$ is
\begin{equation}
\label{lg}
\rho(r,r')=\rho(\mid r-r'\mid)\propto\frac{e^{-\mid r-r'\mid / \xi}}{\mid r-r'\mid}.
\end{equation}
We find that when $r$ approaches $r'$, the density $\rho(\mid r-r'\mid)$ is inversely proportional to $1\mid r-r'\mid$, which means that the central density slope of the subhalos is -1. In Fig.1 we show the density profiles of (\ref{phi}) and the NFW.

Simulations and observations show that the galaxy cluster halos also follow the NFW profile. However, commonly galaxy cluster halos are the most massive bounded system, and it may be not suitable to treat galaxy cluster halos as fluctuations from the equilibrium state of larger-scale halos. Besides, there are also many but few fraction main halos obeying the NFW profile at the subcluster scale.

We suggest the LG model still can guide to explain the central cusp of these halos. In the background cosmology the pressure of the matter is always believed to be zero, because the matter is assumed to be composed of non-relativistic ideal gas and its equation of state is $P_{\mathrm{m}}=\frac{k_{\mathrm{B}} T}{m c^{2}} \rho_{\mathrm{m}} c^{2}$, which has equivalently assumed that there is an equilibrium state for the matter in the Universe. Here we accept this assumption and treat these relaxed and isolated main halos as fluctuations from the equilibrium background Universe, then if $h(r,r')=\delta(r-r')$ their cusps all can be obtained by the second approach of the above section. For these halos near the galaxy cluster scale, another evidence is that from Appendix we find that the LG model equivalently assumes that the density field is random and Gaussian with the power spectrum (\ref{P(k)-app}) and
\begin{equation}
\label{p(k)k-2}
P(k)\propto k^{-2}, k\rightarrow k_c ,
\end{equation}
where $k_c$ is large enough to be at the cluster scale but cannot be larger, or else (\ref{short-range}) can not set up and the non-gaussian effects can never be neglected. (\ref{p(k)k-2}) will finally lead to $r^{-1}$ law of density, because in simulations and observations the density field is represented by a set of mass particles with Poisson sampling, and when $r$ is small, $\rho(r)\propto C(r)\propto r^{-(n+3)}$ if $P(k)\propto k^n$ with random density field \citep{mo10}. It should be noted that: (\ref{p(k)k-2}) is a final result predicted by statistical mechanics, which means that the power spectrum should include the nonlinear effects, although at the scales above galaxy cluster the possible non-Gaussian effects of the density field will be neglected to ensure that the power spectrum still can fundamentally describe the density field. In pure dark matter simulations the power spectrum can be calculated directly with suitable box size; while the power of current observations is still limited and we resort to the emulator including the baryonic (AGN) effects. From Fig.2 we can find when $k$ is near the cluster scale, the power spectrum is consistent with (\ref{p(k)k-2}).

\begin{figure}[h]
\centering
\includegraphics[width = 8cm]{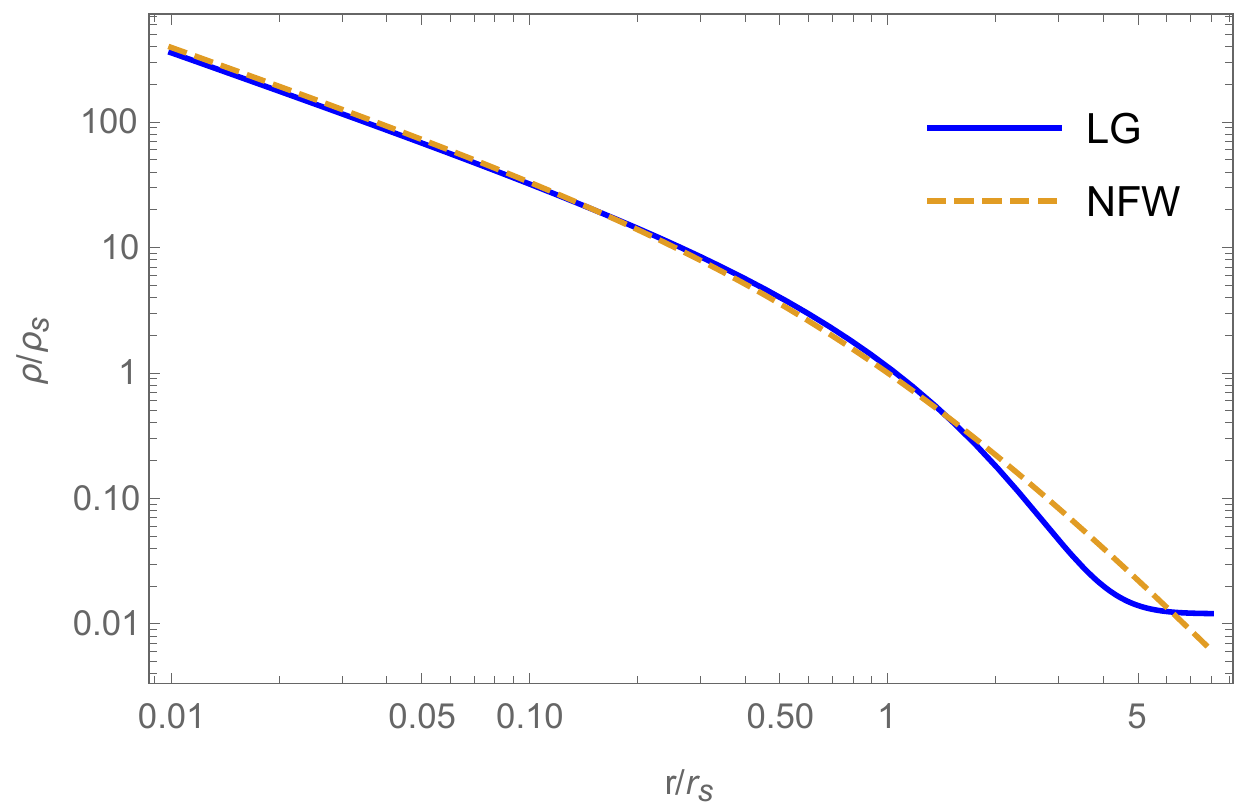}
\caption{The density profile of (\ref{phi}) compared with NFW. $(\ref{phi})$ is parameterized by$\rho_{LG}(r)=\rho_sr_se^{-r/r_s}/r+\rho_0$. In this figure we set $\rho_{LG}(r_s)=\rho_{NFW}(r_s)=\rho_s$ and the concentration $c_{LG}=c_{NFW}=r_{200}/r_s=6$.}
\end{figure}
\begin{figure}[h]
\centering
\includegraphics[width =9.8 cm]{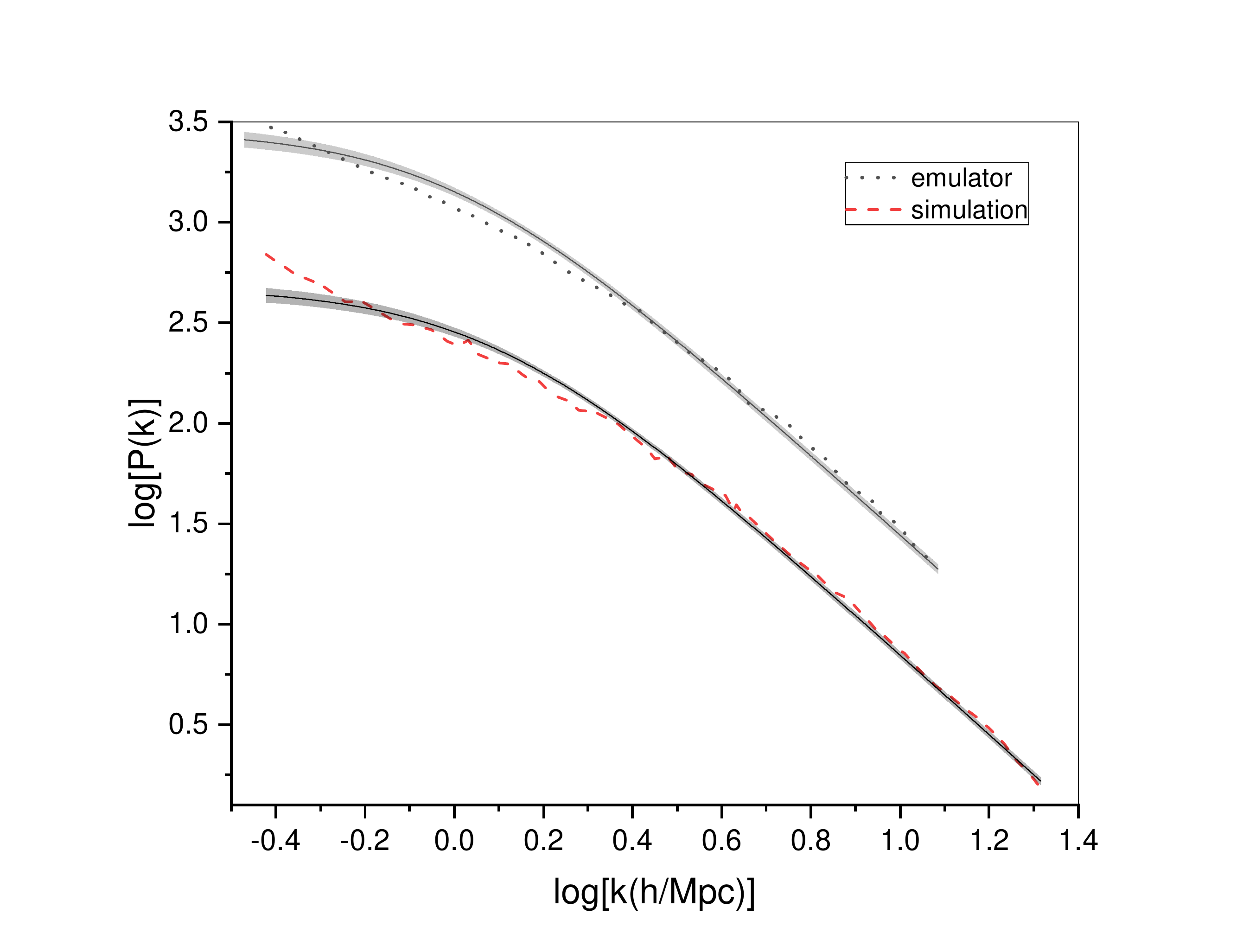}
\caption{The power spectrums in the simulation including the baryonic effects and in emulating the future's observations \cite{heitmann14}, and both of them are fitted by (\ref{P(k)-app}) with 95\% confidencial band.}
\label{fig2}
\end{figure}

\section{Discussion and Conclusion}
The order parameter in LG theory always corresponds to certain symmetries, while the density distribution also can reflect the symmetries of the system, such as that the homogeneous and isotropic systems have translational and rotational invariant symmetries. So, as works in the vapor-liquid interface the order parameter will be the density in this work to explain the cusps of dark matter halos in simulations. As stated in the above section, the subhalos are considered as fluctuations from the equilibrium of the main halos, which, however, may face some arguments, because in CDM models the structure formation of the Universe is bottom-up, i.e. the small structure forms before the large structure. Here it should be emphasized that we just try to explain the coexistent state of the small and large halos after enough time, which does not matter with the time order of the formation of halos, and the substructures also can be destroyed if they can not coexist with the main halos, such as suffering from tidal stripping, merging and others.

Then based on the cosmological principle we suggest the LG model also may provide guidance to reveal the cusp of galaxy cluster halos and other main halos. We think that the background cosmology has equivalently assumed that there is an equilibrium state for all the matters in the Universe. This paper just accepted this assumption and studied the fluctuations from this equilibrium state. The inflated Universe can explain the scale-invariant power spectrum $P(k)\propto k$, which may do not matter with $P(k)=constant$ for $k\rightarrow 0$ (see Appendix) in the LG theory. Here we only focus on the character of $P(k)$ near the galaxy cluster scale, which is shown to be consistent with (\ref{p(k)k-2}) .

From the LG theory, it can be found that the origin of $r^{-1}$ cusp is from the assumption, i.e. the fluctuations of the Helmholtz free energy mainly depend on the density's 'amplitude' and gradient, which may be reasonable for self-gravitating systems, because any macroscopic physical quantities of the self-gravitating system depend on the density distribution which contains the information about the amplitude, the gradient, the second order derivative and others of the density, and for these quantities' fluctuations, the information except the 'amplitude' and gradient may reasonably be neglected above certain space scale, which will be further confirmed. Besides, one problem is that, the $r^{-1}$ cusp in this work does not matter with the nature of dark matter. This problem does not seem to be consistent with these simulations with hot or self-interacting dark matter \citep{maccio12,elbert15} originally proposed to solve the core-cusp problem, and we will further validate these simulations' results and study more general models than this work, such as other forms of the fluctuations of the free energy$\delta F$, other perturbations $h(r)$, the effects of $a_2$ and $b$, etc.


In conclusion, in this paper we consider the cold dark matter halos in simulations as fluctuations from the thermodynamical equilibrium state, and used the LG theory to explain the cusp whose origin still has no consensus.  In the self-gravitating systems, the fluctuations of the Helmholtz free energy will depend mainly on the 'amplitude' and the gradient of the density, which may be the final reason of the almost universality of the $r^{-1}$ cusp. This work also strengthens the point that the equilibrium statistical mechanics may still have a great use for self-gravitating systems. In the future, we will further confirm our assumptions in this work and study the cases of warm dark matter or self-interacting dark matter simulations.



\section{Appendix}
(\ref{correlationlength}) can be obtained as shown in the textbooks such as \citet{zhang05}:
making Fourier expansion of the density contrast,
\begin{equation}
\rho(\textbf{r})-\bar{\rho}=\int\rho_{\textbf{k}} \mathrm{e}^{i \textbf{k} \cdot \textbf{r}}\mathrm{d}^{3} k,
\end{equation}
so
\begin{equation}
\begin{aligned}
\left|\rho_{\textbf{k}}\right|^{2}&=\rho_{\textbf{k}} \rho_{-\textbf{k}}\\
                                  &=\int d^{3} r \int d^{3} \textbf{r'}^{\prime}[\rho(\textbf{r})-\bar{\rho}]\left[\rho\left(\textbf{r}^{\prime}\right)-\bar{\rho}\right] e^{-i k(\textbf{r}-\textbf{r}')}
\end{aligned}
\end{equation}
Ensemble averaging both sides of it,
\begin{equation}
\label{power-corre}
P(k)=\overline{\left|\rho_{k}\right|^{2}}=\int d^{3} r \int d^{3} r^{\prime} C\left(\left|\textbf{r}-\textbf{r}^{\prime}\right|\right) e^{-i\textbf{k}\cdot\left(\textbf{r}-\textbf{r}^{\prime}\right)}
\end{equation}
Then we will calculate $P(k)$ by the LG model (\ref{freeenery}):
\begin{equation}
(\rho(\textbf{r})-\bar{\rho})^{2}= \int\mathrm{d}^{3} k\int\mathrm{d}^{3} k'\rho_{\textbf{k}}^{*} \rho_{\textbf{k}^{\prime}} e^{-i\left(\textbf{k}-\textbf{k}^{\prime}\right) \cdot \textbf{r}},
\end{equation}
\begin{equation}
(\nabla \rho)^{2}=\int\mathrm{d}^{3} k\int\mathrm{d}^{3} k' \rho_{\textbf{k}}^{*} \rho_{\textbf{k}} \textbf{k} \cdot \textbf{k}^{\prime} \mathrm{e}^{-\mathrm{i}\left(\textbf{k}-\textbf{k}^{\prime}\right) \cdot \textbf{r}}
\end{equation}
\begin{equation}
\int(\rho-\bar{\rho}) d^{3} r=0
\end{equation}
\begin{equation}
\begin{aligned} 
\Delta F &=\int\mathrm{d}^{3} k\int\mathrm{d}^{3} k' \rho_{\textbf{k}}^{*} \rho_{\textbf{k}'}\left(\frac{a_{2}}{2}+\frac{b}{2} \textbf{k} \cdot \textbf{k}^{\prime}\right) \int \mathrm{e}^{-\mathrm{i}(\textbf{k}-\textbf{k}') \cdot \mathrm{r}}{\mathrm{d}}^{3} \textbf{r} \\ &=\int\mathrm{d}^{3} k' \rho_{\textbf{k}}^{*} \rho_{\textbf{k}'}\left(\frac{a_{2}}{2}+\frac{b}{2} \textbf{k} \cdot \textbf{k}^{\prime}\right) \delta_{k, k'} \\ &=\frac{1}{2} \int\mathrm{d}^{3} k\left|\rho_{k}\right|^{2}\left(a_{2}+b k^{2}\right)
\end{aligned}
\end{equation}
The probability distribution of the fluctuations of the Helmholtz free energy with fixed volume V (p.294 of \cite{zhang05}) is
\begin{equation}
\label{dstri-deltaF}
w=w_0e^{-\Delta F/k_BT},
\end{equation}
so
\begin{equation}
 w =w_{0} \exp \left[-\frac{1}{2k_{\mathrm{B}} T} \int\mathrm{d}^{3} k\left(a_{2}+b k^{2}\right)\left|\rho_{k}\right|^{2}\right]
\end{equation}
which shows that density perturbation field is Gaussian with power spectrum
\begin{equation}
\label{P(k)-app}
\begin{aligned}
P(k)&=\frac{\int_{0}^{\infty}\left|\rho_{k}\right|^{2} \exp \left[-\frac{a_{2}+b k^{2}}{2 k_{\mathrm{B}} T}\left|\rho_{k}\right|^{2}\right] d\left|\rho_{k}\right|}{\int_{0}^{\infty} \exp \left[-\frac{a_{2}+b k^{2}}{2  k_{\mathrm{B}} T}\left|\rho_{k}\right|^{2}\right] \mathrm{d}\left|\rho_{k}\right|}\\
    &=\frac{{k_{\mathrm{B}} T}}{a_{2}+b k^{2}}
\end{aligned}
\end{equation}
Finally by (\ref{power-corre})
\begin{equation}
C(r)=\frac{k_{\mathrm{B}} T}{(2 \pi)^{3}} \int \frac{\mathrm{e}^{\mathrm{i} \textbf{k} \cdot \mathrm{\textbf{r}}}}{a_{2}+b k^{2}} \mathrm{d}^{3} k=\frac{k_BT}{4\pi b}\frac{e^{-\frac{r}{\xi}}}{r}.
\end{equation}
where the last equality just is a mathematical problem which can be solved by software such as Mathematica.

\end{document}